\documentclass[conference]{IEEEtran}

\usepackage[dvips]{color}
\usepackage{epsf}
\usepackage{times}
\usepackage{epsfig}
\usepackage{amsmath}
\usepackage{amssymb}
\usepackage{amsxtra}
\usepackage{here}
\usepackage{rawfonts}
\usepackage{times}
\usepackage{url}
\usepackage{cite}
\usepackage{float}
\usepackage{subcaption}
\usepackage{graphicx}
\usepackage{color}
\usepackage{booktabs}
\usepackage{multirow}
\usepackage{lipsum}
\usepackage{adjustbox}

\clearpage
\newlength{\aligntop}
\newlength{\alignbot}
\makeatletter

\IEEEoverridecommandlockouts
\begin{document}
\title{\huge Review\! of\! Cyber-Physical\! Attacks\! and\! Counter\! Defense\! Mechanisms\!
for\! Advanced\! Metering\! Infrastructure\! in\! Smart\! Grid}

\author{\IEEEauthorblockN{Longfei Wei, Luis Puche Rondon, Amir Moghadasi, and Arif I. Sarwat}
\IEEEauthorblockA{Department of Electrical and Computer Engineering, Florida International University, Miami, Florida, USA,\\
Emails: {\{lwei004, lpuch002, amogh004, asarwat\}@fiu.edu}}
}


\maketitle

\begin{abstract}
The Advanced Metering Infrastructure (AMI) is a vital element in the current development of the smart grid. AMI technologies provide electric utilities with an effective way of continuous monitoring and remote control of smart grid components. However, owing to its increasing scale and cyber-physical nature, the AMI has been faced with security threats in both cyber and physical domains.
This paper provides a comprehensive review of the crucial cyber-physical attacks and counter defense mechanisms in the AMI. First, two attack surfaces are surveyed in the AMI including the communication network and smart meters. The potential cyber-physical attacks are then reviewed for each attack surface. Next, the attack models and their cyber and physical impacts on the smart grid are studied for comparison. Counter defense mechanisms that help mitigate these security threats are discussed. Finally, several mathematical tools which may help in analysis and implementation of security solutions are summarized.

\end{abstract}

\IEEEpeerreviewmaketitle
\vspace{-0.2cm}
\section{Introduction}\label{sec:intro}\vspace{-0.1cm}
The smart grid is a complex cyber-physical system (CPS) incorporating various spatially distributed subsystems including sensors, actuators, and controllers, which is expected to be a critical technological infrastructure for our nation~\cite{GTD8}. The AMI is one important element of the smart grid being implemented allowing for bi-directional communication between electric utility companies and customers~\cite{AMI3,AMI4,AMI14}. As shown in Fig. 1, the AMI mainly integrates 
the information/communication network, smart meters, and meter data management system (MDMS). The communication network of the AMI is primarily comprised of three important areas including home area network (HAN), wide area network (WAN), and the utility system. Smart meters are the main customer-side installed electronic devices in the AMI, which forward the customers' electricity consumption information to the electric utility. The utility then integrates these information to generate electricity bills, enable demand response, predict user electricity consumption patterns, and update pricing in real time. 


Due to its increasing scale and cyber-physical nature, the security issues of the AMI have become a recent area of interest~\cite{Smart2,AMI19,AMI25,AMI21,AMI22,AMI23,AMI24}. According to the 2014 McAfee report~\cite{AMI25}, 80\% of the surveyed electric utilities have faced at least one large-scale Denial of Service (DoS) attack to their communication networks, and 85\% of the utilities have suffered network infiltrations. Potential cyber security threats and vulnerabilities existing in the AMI are analyzed in~\cite{AMI21}. For a specific DoS attack targeted at the AMI communication network, the attack model and its physical impact are proposed in~\cite{AMI22}, where an attacker may compromise the AMI devices and disrupt data traffic on the network. A malicious attack targetting smart meters is introduced in~\cite{AMI23}, where an attacker can alter the meter measurement data. A cyber attack scenario of an attacker hacking the AMI communication network and performing DoS attacks is simulated in~\cite{AMI24}. 

\begin{figure}[t]
    \label{fig:AMI}
    \includegraphics[width=0.5\textwidth]{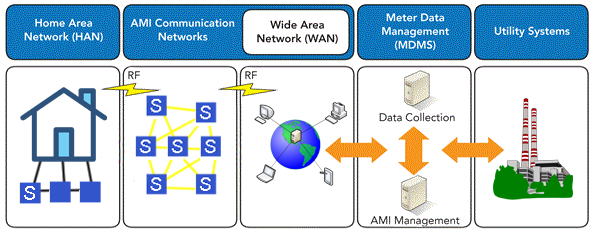}
    \centering
    \caption{The AMI network model and its main components.}
    \vspace{-0.6cm}
\end{figure}

This paper intends to provide a systematic summary of the critical cyber-physical attacks and counter defense mechanisms in the AMI, and survey main mathematical tools for analyzing these attack-defense scenarios. In particular, this paper has three main objectives listed as follows: 
\begin{enumerate}
  \item Survey the two main attack surfaces in the AMI and the potential cyber-physical attacks for each attack surface.
  \item Study the attack models and their cyber and physical impacts on the smart grid.
  \item Summary countermeasure approaches and analytical frameworks which can help in analyzing the AMI attack-defense scenarios and devising proper defense strategies.
\end{enumerate} 


\vspace{-0.2cm}
\section{Cyber-Physical Security Threats in AMI}\label{sec:threat}\vspace{-0.1cm}
As a critical point between electric utilities and customers, the security of AMI is an important area for the smart grid monitoring and operation and the customers' privacy. For example, a malicious attacker can manually compromise smart meters and change the meter measurements, affecting the integrity of reported data. Moreover, the information and communication technologies (ICTs) integrated with the AMI opens window for potential hackers, where cyber attacks can compromise electronic devices, and insert bad data into the communication network. Owing to the expansive deployment of the AMI devices, these cyber-physical attacks can have the potential to disconnect electricity from end consumers, even leading the cascading failures in smart grid and other connected critical infrastructures such as transportation and telecommunications.
In this section, we will review two main attack surfaces in the AMI including the smart meter and the communication network, and provide a summary of the potential cyber-physical attacks targeted at each one.

\begin{figure}[t]
    {\label{fig:AMI}}
    \includegraphics[width=0.5\textwidth]{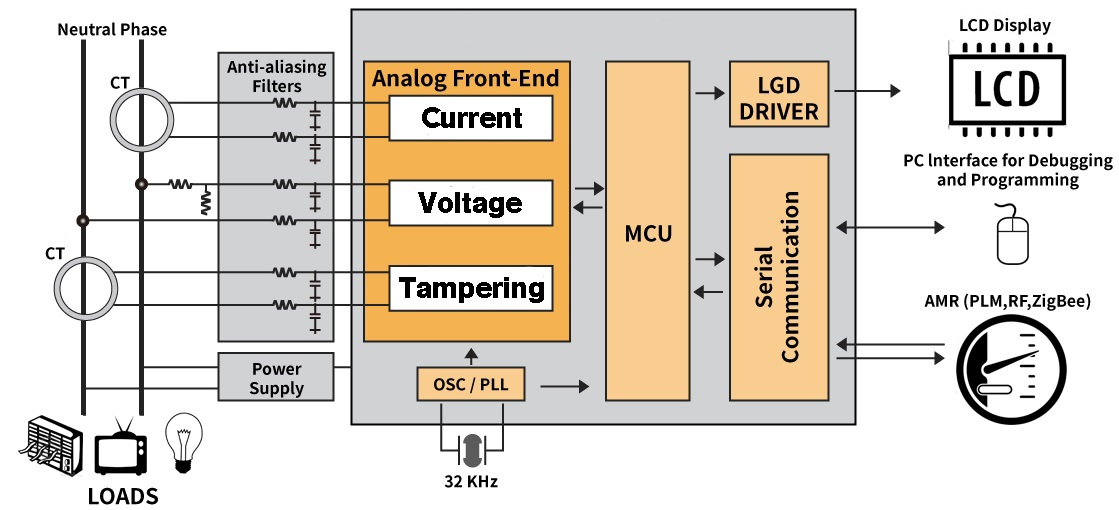}
    \centering
    \caption{Typical elements of smart meter.}
    \vspace{-0.6cm}
\end{figure}
\vspace{-0.2cm}
\subsection{Smart Meters}\label{sec:meter}\vspace{-0.1cm}
Smart meters are electronic devices which records consumption and then reports this information back to the utility, often in assigned intervals. Smart meter facilitates the dash boarding of smart grid system monitoring, automated operation, system recovery, dynamic electricity pricing, and more consumption-based customer services, whose typical elements are shown in Fig. 2. Traditional meters were already susceptible to physical attacks due to their importance, but smart meters open another window that cyber attackers can get access to. In design, smart meters are purchased in bulk (by the millions) and thus driven by low cost. As a result, internal hardware and firmware may be limited. In the sense of capabilities, it means that security often takes a backseat when design must meet both the cost and requirements. Coupled with the vase number in deployment and limited defense resource, a series of the theoretical and demonstrated attacks aimed at compromising smart meters~\cite{AMI5,AMI6,AMI7,AMI8,AMI9,AMI10,AMI11,AMI12,AMI13} are shown as follows:
\begin{itemize}
\item \emph{Denial of Service (DoS) Attacks} compromise smart meters by overwhelming a network or tampering with the routing. This attack can render a meter incapable of responding to any request from electric utilities or consumers. 
\item \emph{False Data Injection Attacks (FDIAs)} insert random and/or deliberate errors within normal smart meter traffic activity to cause corrupted measurements to deliverately cause issues in the smart grid network.
\item \emph{De-pseudonymization Attacks} compromise identity and privacy of smart meter data.
\item \emph{Man-in-the-Middle Attacks} where attackers can place themselves between electric utilities and customers.
\item \emph{Meter Spoofing and Energy Fraud Attacks} can get the ID number of smart meter through physical access.
\item \emph{Authentication Attacks} can authenticate hackers as a valid customer via methods such as stealing a session or acquiring the authentication from memory.
\item \emph{Disaggregation Attacks} attempt to profile customer energy consumption behavior.
\end{itemize}

\begin{table*}[tbp]
\centering
\caption{Cyber and Physical Attacks Targeted at the AMI}
\label{T4}
\begin{tabular}{@{}clll@{}}
\toprule
\multicolumn{2}{c}{\multirow{2}{*}{\textbf{Attack Type}}}                                            & \multicolumn{2}{c}{\textbf{Attack Target}}                                                                                                                                                                                                                                                           \\ \cmidrule(l){3-4} 
\multicolumn{2}{c}{}                                                                                 & \multicolumn{1}{c|}{\textbf{Smart Meter}}                                                                                                                                      & \multicolumn{1}{c}{\textbf{AMI Communication Network}}                                                              \\ \midrule
\multicolumn{2}{c|}{\textbf{Physical}}                                                               & \multicolumn{1}{l|}{\begin{tabular}[c]{@{}l@{}}1. Meter Manipulation\\ 2. Meter Spoofing and Energy Fraud Attack\end{tabular}}                                                    & 1. Physical Attack                                                                                                  \\ \midrule
\multicolumn{1}{c|}{\multirow{3}{*}{\textbf{Cyber}}} & \multicolumn{1}{l|}{\textbf{Availability}}    & \multicolumn{1}{l|}{1. Denial of Service (DoS)}                                                                                                                                & 1. Distributed Denial of Service (DDoS)                                                                             \\ \cmidrule(l){2-4} 
\multicolumn{1}{c|}{}                                & \multicolumn{1}{l|}{\textbf{Integrity}}       & \multicolumn{1}{l|}{1. False Data Injection Attack (FDIA)}                                                                                                                     & 2. False Data Injection Attack (FDIA)                                                                               \\ \cmidrule(l){2-4} 
\multicolumn{1}{c|}{}                                & \multicolumn{1}{l|}{\textbf{Confidentiality}} & \multicolumn{1}{l|}{\begin{tabular}[c]{@{}l@{}}1. De-pseudonymization Attack\\ 2. Man-in-the-middle Attack\\ 3. Authentication Attack\\ 4. Disaggregation Attack\end{tabular}} & \begin{tabular}[c]{@{}l@{}}1. WiFi/ZigBee Attack\\ 2. Internet Attack\\ 3. Data Confidentiality Attack\end{tabular} \\ \bottomrule
\end{tabular}
\vspace{-0.4cm}
\end{table*}

\begin{figure}[t]
    {\label{fig:AMI}}
    \includegraphics[width=0.5\textwidth]{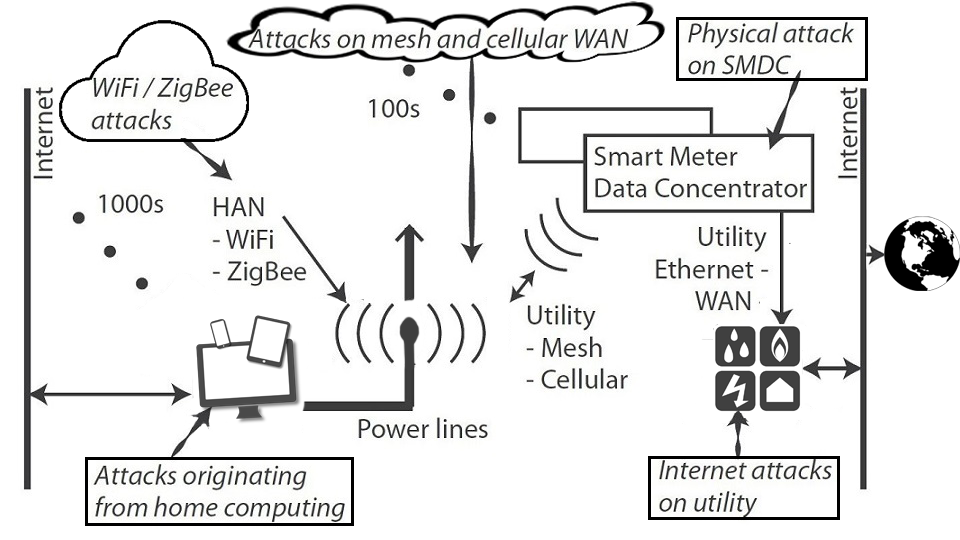}
    \centering
    \caption{The communication network of AMI and the potential cyber-physical attacks targeted at the network.}
    \vspace{-0.6cm}
\end{figure}

\vspace{-0.2cm}
\subsection{AMI Communication Network}\label{sec:infor}\vspace{-0.1cm}
The communication network is a key component of the AMI that links the devices using a wireless Frequency Hopping Spread Spectrum (FHSS) mesh or a similar type cellular network. The AMI communication network usually accomplishes the link to the local HAN on the consumers' side through WiFi, Zigbee or Z-wave protocols. The communication network then connect to the utility’s WAN, which is usually an Ethernet infrastructure. Moreover, the communications network is distributed through an urban sector in company with the smart grid. The scale of this network can vary from a couple of hundreds to thousands of smart meter data collector devices. Each collector is capable of serving thousands of smart meters, raising the number of devices to multiple thousands or millions in total. Therefore, the vulnerabilities of the AMI communication network can be exploited or disabled by attacks on the underlying communication infrastructure, insertion of false user requests, unauthorized alteration of demand side schedules and illegal market manipulation; all of which can impact system operations and result in both power shortage, loss of trust and negative economic impacts. As illustrated in Fig. 3, the potential and demonstrated attacks aimed at the communication network~\cite{AMI16,AMI17,AMI18,AMI19,AMI21,AMI22,AMI23,AMI24,AMI25} are shown as follows:

\begin{itemize}
\item \emph{Distributed Denial of Service (DDoS)} Attacks which target AMI communication networks' data collector, preventing the normal communication between Wide Area Network (WAN) and Neighborhood Area Network (NAN).
\item \emph{False Data Injection Attacks (FDIAs)} introduce random and corrupted data within standard traffic activity in order to cause invalid measurements with the goal of disrupting the AMI network.
\item \emph{Physical Attacks} that compromise the smart meter data collectors, and disrupt the communication between the electric utility and the end customer of power.
\item \emph{WiFi/ZigBee Attacks} in the AMI communication networks that attack the WiFi/ZigBee networks in Home Area Network (HAN).
\item \emph{Internet Attacks} that compromise the software and systems in electric utilities.
\item \emph{Data Confidentiality Attack} attempts to compromise the information between electric utilities and end customers by targeting the hardware within the AMI communication network.
\end{itemize}

A summary of the different types of cyber and physical attacks targeted at the AMI systems including the smart meter and the communication network is shown in Table {\ref{T4}}.

\section{Defense Mechanisms in Smart Meters}\label{sec:smartmeter}
Among the cyber-physical security threats faced by the smart meters, \emph{electricity theft} is a major security challenge caused by meter manipulation and FDIA, where malicious attackers can alter consumption measurements collected by smart meters. According to a World Bank report~\cite{Report6,GTD6}, electricity theft reaches up to 50\% in some jurisdictions of developing countries. The significance of security for smart meters has been a well-researched topic in the literature, where the focus is on ensuring power availability at all times~\cite{AMI6,GTD7}. \emph{Traditional research} for detecting electricity theft has focused on implementing specific devices, like wireless sensors and balance meters, to provide a high theft detection rate~\cite{AMI17,AMI18}. An anti-tampering sensor based AMI intrusion detection mechanism was introduced in~\cite{AMI17}, where anti-tampering sensors were embedded into smart meters. In~\cite{AMI18}, a limited number of balanced meters were installed in the smart grid distribution network, so that the system operator can detect whether abnormal smart meters existing in the network. Since additional devices need to be installed in these works, the cost of detecting the abnormal actions of millions of smart meters will be significantly increases. Moreover, all of these works cannot detect specific smart meters that being hacked.

\vspace{-0.2cm}
\subsection{Machine Learning for Electricity Theft Detection}\label{sec:machtheft}\vspace{-0.1cm}

Recently, machine learning have been used to train a classifier based on detailed electricity usage measurements, which aims to classify the normal usage versus electricity theft. The basic procedure of this approach consists of seven parts: data collection and preprocessing, feature extraction, machine learning based classifier training, data classification, and suspected electricity theft generation.

In~\cite{AMI9}, Principal Component Analysis (PCA) based theft detection was proposed to detect abnormalities in electricity consumption behavior. A method is proposed which leverages the Density-Based Spatial Clustering of Applications with Noise (DBSCAN) algorithm, and this procedure is shown to properly detect abnormalities in consumption behavior when used with PCA. There are three advantages with the use of this method. First, the researchers could calculate the consumption trends which repeat over time by extracting principal components which retain the maximum amount of variance in the data. Components that belong in the lower variance are filtered out as ‘noise’ in usage behavior. Second, to process the massive amount of data, the two principal components are noted to allow visualization in a two-dimensional space. Third, anomaly detection takes place in a set of data that includes the usage of all consumers. This means that an attacker would have trouble reverse engineering and avoiding detection as it would require two things. Complete knowledge of every smart meter and all consumer information being within the communications network. 

In~\cite{AMI10}, usage data was proved to be non-stationary and Auto Regressive Integrated Moving Average (ARIMA) forecasting methods were proposed to validate readings. First, the ARMA model is ill-suited for anomaly detection in electricity consumption since most customers use up their power in a non-stationary manner. The ARIMA forecasting methods are introduced for validating electricity consumption readings. Second, to evaluate the effectiveness of forecasting with ARIMA; a scenario where smart meters were tampered with to allow electricity theft was evaluated. Third, introducing complementary checks on factors such as mean and variance was found to aid in the mitigation of electricity theft by 77.56\%

In~\cite{AMI11}, CPBETF (Consumption Pattern-Based Energy Theft Detector) which employs a multi-class Support Vector Machine (SVM) for each customer was formulated. In this detection, transformer meters measure the total consumption of each neighborhood. Then, the total usage reported by the smart meters is compared to the measurements from the transformers. Users and clients will be marked as suspicious if a nontechnical loss (NTL) is detected at this level. A User’s past data as well as synthetic attack datasets are used to train a multi-class SVM (support vector machine). With this, a classifier can be generated to determine whether the recent sample is either benign or malicious.

Apart from the above techniques many other classification methods exist for detection of energy theft. Methods such as, fuzzy logic classification and neural networks are also feasible. However, these works ignore the attack models of potential thieves and the effectiveness of anomaly detector was only evaluated based on given datasets of attack examples.

\begin{table*}[tbp]
\centering
\caption{Different Application Techniques for Electricity Theft Detection}
\label{T5}
\begin{adjustbox}{max width=\textwidth}
\begin{tabular}{@{}l|l|l|l@{}}
\toprule
\multicolumn{2}{l|}{\textbf{Electricity Theft Detection Techniques}} & \textbf{Advantage} & \textbf{Disadvantage} \\ \midrule
\multirow{2}{*}{ \begin{tabular}[c]{@{}l@{}} \textbf{Traditional Method} \\ {[}22-26{]} \end{tabular}} & \begin{tabular}[c]{@{}l@{}}Anti-tampering Sensor {[}22{]}\end{tabular} & \multirow{2}{*}{1. Reduce risks due to non-billed electricity} & \multirow{2}{*}{\begin{tabular}[c]{@{}l@{}}1. Not identify specific meters being compromised\\ 2. Increase the cost of deploying and operating\end{tabular}}   \\ \cmidrule(lr){2-2}
& \begin{tabular}[c]{@{}l@{}} Trusted Balanced Meter {[}23{]}\end{tabular}  & & \\ \midrule
\multirow{3}{*}{\begin{tabular}[c]{@{}l@{}} \textbf{Machine Learning} \\ {[}16-18{]} \end{tabular}}   & \begin{tabular}[c]{@{}l@{}}PCA based detection  {[}16{]}\end{tabular}                & \multirow{3}{*}{\begin{tabular}[c]{@{}l@{}}1. Implement of Meter Data Management System\\ 2. Train a classifier based on usage data\\ 3. Increase effectiveness of theft detection\end{tabular}} & \multirow{3}{*}{\begin{tabular}[c]{@{}l@{}}1. Ignore the attack models of potential thieves \\ 2. Evaluate effectiveness only on given attack example\\ 3. Ignore the privacy of customer consumption \end{tabular}} \\ \cmidrule(lr){2-2}
& \begin{tabular}[c]{@{}l@{}} ARIMA based Detection {[}17{]}\end{tabular} &  \\ \cmidrule(lr){2-2}
& \begin{tabular}[c]{@{}l@{}} SVM based Detection {[}18{]}\end{tabular} &  &  \\ \midrule
\textbf{Game Theory} {[}12{]} & \begin{tabular}[c]{@{}l@{}}Stackelberg Game {[}12{]}\end{tabular}                & \begin{tabular}[c]{@{}l@{}}1. Propose interactions between utility and thieves\\ 2. Consider the worst case of electricity theft \\ 3. Consider the privacy of customer consumption \end{tabular}          &  1. Ignore coordination between the electricity thieves \\ \bottomrule
\end{tabular}
\end{adjustbox}
\vspace{-0.4cm}
\end{table*}

\vspace{-0.2cm}
\subsection{Game Theory for Electricity Theft Detection}\label{sec:gametheft}\vspace{-0.1cm}

Game theoretic theft detection schemes have been proposed recently and provide another angle on solving the electricity theft issues~\cite{AMI12,AMI5}. To solve the problem of theft detection, a single leader, multi-follower {Stackelberg game} can be formulated between the utility and thieves to characterize strategic interactions between the two~\cite{AMI5}. In this game, the utility intended to maximize the detection probability and minimize the investment in monitoring fraud. On the other hand, each electricity thief was to steal a certain amount of power and minimize the probability being uncovered. Based on the Nash equilibrium of the formulated game, the optimal strategies for the defender are derived for selecting the sample rate and the optimal tariff.

However, these works assume all electricity thieves as a player, and the competition between thieves was ignored in the model. If thieves add high loads to the distribution networks and steal electricity at the same time, the resulting power surges and electrical system failures can cause power outages, raising the possibility of thefts being detected. A summary of the different applications for electricity theft in the AMI is shown in Table~\ref{T5}.

\section{Defense Mechanisms in AMI Communication Network}\label{sec:network}

In this section, we first study the model of the AMI communication network. Then, we survey two specific game-theoretic models for defending the communication network against \emph{data confidentiality attacks} and  \emph{distributed DoS attacks}, respectively. 

Consider the AMI communication network as a tree-pattern architecture $\mathcal{T}$ with one root node, where nodes represent the AMI devices. Let $\mathcal{N} = \{1, 2,...,N\}$ denote the set of 
nodes in $\mathcal{T}$, where $N$ is the total number of nodes, and the root node is referred as $1$.
Each node $i\in{\mathcal{N}\setminus\{1\}}$ records data from its children nodes $Ch(i)$, combines this data as a whole, and finally sends it to its respective parent node $f(i)$. Assume that a total of $N$ aggregation levels existing in $\mathcal{T}$, and let $N_i$ denote the set of nodes belonging to the $i$-th aggregation level. In the attack-defense scenario, the attacker can select each node for attacking. Therefore, game theory provides a way for the defender to find the optimal defense mechanism.

\vspace{-0.2cm}
\subsection{Defense Mechanisms for Data Confidentiality Attacks}\label{sec:confident}\vspace{-0.1cm}

\begin{table}[tbp]
\vspace{0.2cm}
\centering
\caption{\small{Game Theoretic Applications}}
\label{T6}
\begin{adjustbox}{max width=0.45\textwidth}
\begin{tabular}{@{}ll@{}}
\toprule
\textbf{Game Theoretic Applicaitons}                                                                                          & \textbf{Game Features}                                                                                                    \\ \midrule
\begin{tabular}[c]{@{}l@{}}Defending AMI communication network \\ against data confidentiality attacks {[}28{]}\end{tabular}  & \begin{tabular}[c]{@{}l@{}}1. Noncooperative\\ 2. Static\\ 3. Two-player\\ 4. Nash equilibrium\end{tabular}               \\ \midrule
\begin{tabular}[c]{@{}l@{}}Protecting AMI communication network \\ against data confidentiality attacks {[}29{]}\end{tabular} & \begin{tabular}[c]{@{}l@{}}1. Noncooperative\\ 2. Stackerlberg\\ 3. Two-player\\ 4. Stackerlberg equilibrium\end{tabular} \\ \midrule
\begin{tabular}[c]{@{}l@{}}Protecting AMI communication network \\ against DDoS attacks {[}31{]}\end{tabular}                 & \begin{tabular}[c]{@{}l@{}}1. Noncooperative\\ 2. Bayesian\\ 3. Multiple-player\\ 4. Nash equilibrium\end{tabular}        \\ \bottomrule
\end{tabular}
\end{adjustbox}
\vspace{-0.6cm}
\end{table}

In~\cite{AMI1}, the data confidentiality attacks in the AMI communication network are analyzed. In this attack-defense scenario, the attacker aims to compromise the AMI data by attacking the nodes of the communication network $\mathcal{T}$ without being detected. Correspondingly, out of a set, the defender may pick one security mode available for each node. A two-player noncooperative game is formulated to model the interaction between the two. In this game, the attacker's strategy is defined as the probability $p_i$ of attacking node $i$, which is subject to a budget limitation, $\sum_i{p_i}\leq{P}\leq{1},\forall{i}$. In contrast, the defender's strategy is defined as the encryption rate $s_i$ of the packets at node $i$, which is subject to a budget constraint $\sum_i{s_i}\leq{S}\leq{N},\forall{i}$. The attacker and defender's utility functions are determined by the data value or security asset $W_i$ for each node $i$.
Take a case where the defender and attacker know everything about a system. In this case the \emph{Nash equilibrium} by definition is the optimal arrangement which provides the most utility to each player from the actions of other players~\cite{Gamebook2,Gamebook3}. In the end, The Nash equilibrium can derive the behavior to be expected from both attacker and defender.

However, assume the attacker chooses their strategy based on based on security techniques deployed in the target system. Therefore, the interactions between the two can be formulated by a Stackelberg game~\cite{Gamebook5}. In this game, the defender acts as a leader which attempts to compose encryption rates. The defender’s purpose is to adjust encryption rates to protect the security of the most amount of data possible. The method which the Stackelberg games solve the problem is backwards induction, producing a solution known as Stackelberg Equilibrium (SE) ~\cite{Gamebook3}. In this game, the defender may anticipate the attacker’s actions, find an efficient defense budget, and create the optimal encryption rate on each device in the AMI with the help of the SE in order to mitigate attacks. 

\vspace{-0.2cm}
\subsection{Defense Mechanisms for Distributed DoS Attacks}\label{sec:DDoS}\vspace{-0.1cm}

In~\cite{AMI26}, the honeypot based defense mechanism is implemented for countering distributed DoS attacks in the AMI communication network, where honeypots are defined as defense resources that help lure, discover, and gather attack information. A Bayesian honeypot game model is formulated  between benign users and malicious ones. The equilibrium in conditions can be achieved for deriving the strategies in use of honeypots and anti-honeypots.

The Bayesian game is defined as follows: $\mathcal{G}_1$ as $\mathcal{G}_1\triangleq\{\{\mathcal{Z},\mathcal{W}\},\{\mathcal{F}_{\mathcal{Z}},\mathcal{F}_{\mathcal{W}}\},\{\mathcal{J}_{\mathcal{Z}},\mathcal{J}_{\mathcal{W}}\}\}$, where $\mathcal{Z}\triangleq\{{Z}_1,{Z}_2,{Z}_3\}$ usually represent an array of services such as: honeypots, real communications, and anti-honeypots. As provided by the smart grid; $\mathcal{W}\triangleq\{{W}_1,{W}_2\}$ is the set of unique visitors: in this case non-malicious users and malicious attackers. $\{\mathcal{F}_{\mathcal{Z}},\mathcal{F}_{\mathcal{W}}\}$ denotes the set of strategies used by the attackers and honeypots respectively. $\mathcal{F}_{\mathcal{Z}}\triangleq\{\Omega_1,\Omega_2\}$ denotes a binary variable. $\Omega_1$ represents a service which is being provided. $\mathcal{F}_{\mathcal{W}}\triangleq\{\Lambda_1,\Lambda_2\}$ also represent a set of binary variables. To represent providing access: $\Lambda_1$ is used. $\{\mathcal{J}_{\mathcal{Z}},\mathcal{J}_{\mathcal{W}}\}$ denotes player payoff, where $\mathcal{J}_{\mathcal{Z}}$ represents the real server payoffs and $\mathcal{J}_{\mathcal{W}}$ represents the payoff for the visitors. The payoffs of legitimate users and attackers are analyzed via game trees. To evaluate the overall performance of the proposed scheme, an AMI network testbed is constructed. A summary of the different game-theoretic applications for protecting the AMI communication network is shown in Table~\ref{T6}.

\section{Conclusion}\label{sec:conc}
In this paper, we provide a comprehensive overview on the potential cyber and physical attacks targeted at the AMI, especially in smart meters and the communication network. We have identified the main security threat for smart meters: electricity theft, and categorized three detection mechanisms including device implementation, machine learning and game theory. Machine learning provides a more efficient way for theft detection than device implementation. And game theory formulates the interaction model between utility and thieves for optimal detection strategies. Game theory is expected to become a key analysis tool for analyzing cyber-physical security issues. Therefore, for AMI communication network, we survey two specific game-theoretic models for protecting the network against data confidentiality attacks and distributed DoS attacks, respectively. As we have reviewed, game theory provides a way to predict the rational attack actions and derive the optimal defense strategies against potential attacks.

\section{Acknowledgement}
This work was supported by the National Science Foundation under Grants CNS-1553494 (NSF) and 800006104 (DOE).

\bibliographystyle{IEEEtran}
\bibliography{references}

\begin{thebibliography}{10}
\providecommand{\url}[1]{#1}
\csname url@samestyle\endcsname
\providecommand{\newblock}{\relax}
\providecommand{\bibinfo}[2]{#2}
\providecommand{\BIBentrySTDinterwordspacing}{\spaceskip=0pt\relax}
\providecommand{\BIBentryALTinterwordstretchfactor}{4}
\providecommand{\BIBentryALTinterwordspacing}{\spaceskip=\fontdimen2\font plus
\BIBentryALTinterwordstretchfactor\fontdimen3\font minus
  \fontdimen4\font\relax}
\providecommand{\BIBforeignlanguage}[2]{{%
\expandafter\ifx\csname l@#1\endcsname\relax
\typeout{** WARNING: IEEEtran.bst: No hyphenation pattern has been}%
\typeout{** loaded for the language `#1'. Using the pattern for}%
\typeout{** the default language instead.}%
\else
\language=\csname l@#1\endcsname
\fi
#2}}
\providecommand{\BIBdecl}{\relax}
\BIBdecl

\bibitem{GTD8}
L.~Wei, A.~Sarwat, W.~Saad, and S.~Biswas, ``Stochastic games for power grid
  protection against coordinated cyber-physical attacks,'' \emph{IEEE Trans. on
  Smart Grid}, vol.~PP, no.~99, pp. 1--1, 2017.

\bibitem{AMI3}
I.~Parvez, A.~Sundararajan, and A.~Sarwat, ``Frequency band for han and nan
  communication in smart grid,'' in \emph{IEEE Symposium on Computational
  Intelligence Applications in Smart Grid (CIASG)}, Orlando, Dec. 2014.

\bibitem{AMI4}
A.~Anzalchi and A.~Sarwat, ``A survey on security assessment of metering
  infrastructure in smart grid systems,'' in \emph{IEEE Southeast Conference},
  Fort Lauderdale, 2015.

\bibitem{AMI14}
I.~Parvez, A.~I. Sarwat, L.~Wei, and A.~Sundararajan, ``Securing metering
  infrastructure of smart grid: A machine learning and localization based key
  management approach,'' \emph{Energies}, vol.~9, no.~9, 2016.

\bibitem{Smart2}
P.-Y. Chen, S.~M. Cheng, and K.-C. Chen, ``Smart attacks in smart grid
  communication networks,'' \emph{IEEE Commun. Mag.}, vol.~50, no.~8, pp.
  24--29, 2012.

\bibitem{AMI19}
S.~Sridhar, A.~Hahn, and M.~Govindarasu, ``Cyber attack-resilient control for
  smart grid,'' in \emph{2012 IEEE PES Innovative Smart Grid Technologies
  (ISGT)}, Jan 2012, pp. 1--3.

\bibitem{AMI25}
S.~Baker, N.~Filipiak, and K.~Timlin, ``In the dark: Crucial industries
  confront cyber attacks,'' 2014.

\bibitem{AMI21}
E.~Naone, ``Hacking the smart grid,'' 2010.

\bibitem{AMI22}
P.~Yi, T.~Zhu, Q.~Zhang, Y.~Wu, and J.~Li, ``A denial of service attack in
  advanced metering infrastructure network,'' in \emph{2014 IEEE International
  Conference on Communications (ICC)}, June 2014, pp. 1029--1034.

\bibitem{AMI23}
P.~McDaniel and S.~McLaughlin, ``Security and privacy challenges in the smart
  grid,'' \emph{IEEE Security Privacy}, vol.~7, no.~3, pp. 75--77, May 2009.

\bibitem{AMI24}
F.~M. Cleveland, ``Cyber security issues for advanced metering infrasttructure
  (ami),'' in \emph{2008 IEEE Power and Energy Society General Meeting -
  Conversion and Delivery of Electrical Energy in the 21st Century}, July 2008,
  pp. 1--5.

\bibitem{AMI5}
S.~Amin, G.~A. Schwartz, A.~A. Cardenas, and S.~S. Sastry, ``Game-theoretic
  models of electricity theft detection in smart utility networks: Providing
  new capabilities with advanced metering infrastructure,'' \emph{IEEE Control
  Systems}, vol.~35, no.~1, pp. 66--81, Feb 2015.

\bibitem{AMI6}
R.~Jiang, R.~Lu, Y.~Wang, J.~Luo, C.~Shen, and X.~S. Shen, ``Energy-theft
  detection issues for advanced metering infrastructure in smart grid,''
  \emph{Tsinghua Science and Technology}, vol.~19, no.~2, pp. 105--120, April
  2014.

\bibitem{AMI7}
S.~McLaughlin, B.~Holbert, A.~Fawaz, R.~Berthier, and S.~Zonouz, ``A
  multi-sensor energy theft detection framework for advanced metering
  infrastructures,'' \emph{IEEE Journal on Selected Areas in Communications},
  vol.~31, no.~7, pp. 1319--1330, July 2013.

\bibitem{AMI8}
E.~de~Buda, ``System for accurately detecting electricity theft,'' Patent US
  20\,100\,007\,336 A1, January, 2010.

\bibitem{AMI9}
V.~Badrinath~Krishna, G.~A. Weaver, and W.~H. Sanders, \emph{PCA-Based Method
  for Detecting Integrity Attacks on Advanced Metering Infrastructure}.\hskip
  1em plus 0.5em minus 0.4em\relax Cham: Springer International Publishing,
  2015, pp. 70--85.

\bibitem{AMI10}
V.~Badrinath~Krishna, R.~K. Iyer, and W.~H. Sanders, \emph{ARIMA-Based Modeling
  and Validation of Consumption Readings in Power Grids}.\hskip 1em plus 0.5em
  minus 0.4em\relax Cham: Springer International Publishing, 2016, pp.
  199--210.

\bibitem{AMI11}
P.~Jokar, N.~Arianpoo, and V.~C.~M. Leung, ``Electricity theft detection in ami
  using customers' consumption patterns,'' \emph{IEEE Transactions on Smart
  Grid}, vol.~7, no.~1, pp. 216--226, Jan 2016.

\bibitem{AMI12}
A.~A. Cárdenas, S.~Amin, G.~Schwartz, R.~Dong, and S.~Sastry, ``A game theory
  model for electricity theft detection and privacy-aware control in ami
  systems,'' in \emph{2012 50th Annual Allerton Conference on Communication,
  Control, and Computing (Allerton)}, Oct 2012, pp. 1830--1837.

\bibitem{AMI13}
S.~Amin, G.~A. Schwartz, A.~A. Cardenas, and S.~S. Sastry, ``Game-theoretic
  models of electricity theft detection in smart utility networks: Providing
  new capabilities with advanced metering infrastructure,'' \emph{IEEE Control
  Systems}, vol.~35, no.~1, pp. 66--81, Feb 2015.

\bibitem{AMI16}
F.~Milano, C.~Canizares, and M.~Invernizzi, ``Multi-objective optimization for
  pricing system security in electricity markets,'' \emph{IEEE Trans. on Power
  Syst.}, vol.~18, no.~2, pp. 596--604, 2003.

\bibitem{AMI17}
K.~Xie, Y.-H. Song, J.~Stonham, E.~Yu, and G.~Liu, ``Decomposition model and
  interior point methods for optimal spot pricing of electricity in
  deregulation environments,'' \emph{IEEE Trans. Power Syst.}, vol.~15, no.~1,
  pp. 39--50, 2000.

\bibitem{AMI18}
M.~Esmalifalak, G.~Shi, Z.~Han, and L.~Song, ``Bad data injection attack and
  defense in electricity market using game theory study,'' \emph{IEEE Trans.
  Smart Grid}, vol.~4, no.~1, pp. 160--169, Mar. 2013.

\bibitem{Report6}
``{EBIOS} expression of needs and identification of security objectives risk
  management method,,'' ANSSI, Tech. Rep., January 2010.

\bibitem{GTD6}
L.~Wei, A.~H. Moghadasi, A.~Sundararajan, and A.~Sarwat, ``Defending mechanisms
  for protecting power systems against intelligent attacks,'' in \emph{Proc.
  IEEE 10th SoSE Conf.}, San Antonio, the United States, May 2015.

\bibitem{GTD7}
L.~Wei, A.~I. Sarwat, and W.~Saad, ``Risk assessment of coordinated
  cyber-physical attacks against power grids: A stochastic game approach,'' in
  \emph{2016 IEEE Industry Applications Society Annual Meeting}, Oct 2016, pp.
  1--7.

\bibitem{AMI1}
Z.~Ismail, J.~Leneutre, D.~Bateman, and L.~Chen, ``A game theoretical analysis
  of data confidentiality attacks on smart-grid ami,'' \emph{IEEE Journal on
  Selected Areas in Communications}, vol.~32, no.~7, pp. 1486--1499, July 2014.

\bibitem{Gamebook2}
R.~B. Myerson, \emph{Game Theory, Analysis of Conflict}.\hskip 1em plus 0.5em
  minus 0.4em\relax Cambridge, MA, USA: Harvard University Press, Sep. 1991.

\bibitem{Gamebook3}
L.~Shapley, ``Stochastic games,'' in \emph{Proc. Nat. Acad. Sci. USA}, vol.~39,
  1953, pp. 1095--1100.

\bibitem{Gamebook5}
A.~Neyman and S.~Sorin, \emph{Stochastic Games and Applications}.\hskip 1em
  plus 0.5em minus 0.4em\relax New York: Kluwer Academic, Jul. 1999.

\bibitem{AMI26}
K.~Wang, M.~Du, S.~Maharjan, and Y.~Sun, ``Strategic honeypot game model for
  distributed denial of service attacks in the smart grid,'' \emph{IEEE
  Transactions on Smart Grid}, vol.~PP, no.~99, pp. 1--1, 2017.

\end{thebibliography}
\end{document}